\documentclass[apj]{emulateapj}

\usepackage{hyperref}
\usepackage{textcomp}
\usepackage{bm,color}
\usepackage{verbatim}
\usepackage{amsmath}
\usepackage{ulem}
\newcommand{\edit}[1]{{#1}}

\newcommand{\pb}{\textsc{Polarbear}}

\newcommand{\sptpol}{{\sc SPTpol}}

\newcommand{\actpol}{{\sc actpol}}

\newcommand{\cmb}{CMB}

\newcommand{\bicep}{BICEP}

\newcommand{\mc}{MC}
\newcommand{\lcdm}{$\Lambda$CDM}
\newcommand{\keckarray}{\textit{Keck Array}}

\newcommand{\clbb}{$C_\ell^{BB}$}

\newcommand{\Alens}{$A_\mathrm{L}$}

\newcommand{\alensvalanderr}{$1.33 \pm 0.32\ \textrm{(statistical)} \pm 0.02\ \textrm{(systematic)} \pm 0.07\ \textrm{(foreground)}$}
\newcommand{\alensval}{$1.33$}

\newcommand{\lenssig}{$4.1\sigma$}
\newcommand{\nolenssig}{$10.9\sigma$}
\newcommand{\Alense}{A_\mathrm{L}}
\newcommand{\sigmaAlens}{$\sigma_{A}$}

\newcommand{\Alenscont}{$A_\mathrm{L}^c$}
\newcommand{\sigmaAlenscont}{$\sigma_{A}^c$}
\newcommand{\deltaAlenscont}{$\Delta A_\mathrm{L}^c$}
\newcommand{\Alensconte}{A_\mathrm{L}^c}
\newcommand{\sigmaAlensconte}{\sigma_{A}^c}

\newcommand{\beq}{\begin{equation}}
\newcommand{\eeq}{\end{equation}}
\newcommand{\bea}{\begin{eqnarray}}
\newcommand{\eea}{\end{eqnarray}}

\newcommand{\Ell}{{\bm\ell}}
\newcommand{\LL}{{\bm L}}
\newcommand{\spp}{{PB17}}
\newcommand{\ipplensing}{{PB14}}
\newcommand{\chinullb}{$\chi_{\rm null}(b)$}
\newcommand{\chisqnullb}{$\chi^2_{\rm null}(b)$}

\newcommand{\chisqnull}{$\chi^2_{\rm null}$}

\newcommand{\cov}{\mathcal{C}_{L\alpha,L'\alpha'}}
\newcommand{\invcov}{\mathcal{C}^{-1}_{L\alpha,L'\alpha'}}

\newcommand{\hsc}{Subaru Hyper Suprime-Cam}
\newcommand{\herschel}{\textit{Herschel}-ATLAS}

\begin{document}
%%%%%%%%%%%%%%%%%%%%% Title, etc. %%%%%%%%%%%%%%%%%%%%%
\title{Measurement of the Cosmic Microwave Background Polarization Lensing Power Spectrum from Two Years of POLARBEAR Data}

%\author{The \pb\ Collaboration}\noaffiliation

\author{The \pb\ Collaboration:
M.~Aguilar Fa\'undez\altaffilmark{5,3}, % fixed
K.~Arnold\altaffilmark{11},
C.~Baccigalupi\altaffilmark{16,14,20},
D.~Barron\altaffilmark{6},
D.~Beck\altaffilmark{1},
S.~Beckman\altaffilmark{10},
F.~Bianchini\altaffilmark{25},
J.~Carron\altaffilmark{4},
K.~Cheung\altaffilmark{10},
Y.~Chinone\altaffilmark{9,18},
H.~El Bouhargani\altaffilmark{1}, % fixed
T.~Elleflot\altaffilmark{11},
J.~Errard\altaffilmark{1},
G.~Fabbian\altaffilmark{4},
C.~Feng\altaffilmark{12},
T.~Fujino\altaffilmark{27},
N.~Goeckner-Wald\altaffilmark{10,8},
T.~Hamada\altaffilmark{2},
M.~Hasegawa\altaffilmark{13,24},
M.~Hazumi\altaffilmark{13,15,18,24},
C.A.~Hill\altaffilmark{10,21},
H.~Hirose\altaffilmark{27},
O.~Jeong\altaffilmark{10},
N.~Katayama\altaffilmark{18},
B.~Keating\altaffilmark{11},
S.~Kikuchi\altaffilmark{27},
A.~Kusaka\altaffilmark{21,9,19,23},
A.T.~Lee\altaffilmark{10,21,22},
D.~Leon\altaffilmark{11},
E.~Linder\altaffilmark{26,21},
L.N.~Lowry\altaffilmark{11},
F.~Matsuda\altaffilmark{18},
T.~Matsumura\altaffilmark{18},
Y.~Minami\altaffilmark{13},
M.~Navaroli\altaffilmark{11},
H.~Nishino\altaffilmark{23},
A.T.P.~Pham\altaffilmark{25},
D.~Poletti\altaffilmark{16},
G.~Puglisi\altaffilmark{8},
C.L.~Reichardt\altaffilmark{25},
Y.~Segawa\altaffilmark{24,13},
B.D.~Sherwin\altaffilmark{17},
M.~Silva-Feaver\altaffilmark{11},
P.~Siritanasak\altaffilmark{11},
R.~Stompor\altaffilmark{1},
A.~Suzuki\altaffilmark{21},
O.~Tajima\altaffilmark{7},
S.~Takatori\altaffilmark{24,13},
D.~Tanabe\altaffilmark{24,13},
G.P.~Teply\altaffilmark{11},
C.~Tsai\altaffilmark{11}
}

\altaffiltext{1}{AstroParticule et Cosmologie (APC), Univ Paris Diderot, CNRS/IN2P3, CEA/Irfu, Obs de Paris, Sorbonne Paris Cit\'e, France}
\altaffiltext{2}{Astronomical Institute, Graduate School of Science, Tohoku University, Sendai, 980-8578, Japan}
\altaffiltext{3}{Departamento de F\'isica, FCFM, Universidad de Chile, Blanco Encalada 2008, Santiago, Chile}
\altaffiltext{4}{Department of Physics \& Astronomy, University of Sussex, Brighton BN1 9QH, UK}
\altaffiltext{5}{Department of Physics and Astronomy, Johns Hopkins University, Baltimore, MD 21218, USA}
\altaffiltext{6}{Department of Physics and Astronomy, University of New Mexico, Albuquerque, New Mexico, USA 87131}
\altaffiltext{7}{Department of Physics, Kyoto University, Kyoto 606-8502, Japan}
\altaffiltext{8}{Department of Physics, Stanford University, Stanford, CA, 94305}
\altaffiltext{9}{Department of Physics, The University of Tokyo, Tokyo 113-0033, Japan}
\altaffiltext{10}{Department of Physics, University of California, Berkeley, CA 94720, USA}
\altaffiltext{11}{Department of Physics, University of California, San Diego, CA 92093-0424, USA}
\altaffiltext{12}{Department of Physics, University of Illinois at Urbana-Champaign, 1110 W Green St, Urbana, IL, 61801, USA}
\altaffiltext{13}{High Energy Accelerator Research Organization (KEK), Tsukuba, Ibaraki 305-0801, Japan}
\altaffiltext{14}{Institute for Fundamental Physics of the Universe (IFPU), Via Beirut 2, 34151, Grignano (TS), Italy}
\altaffiltext{15}{Institute of Space and Astronautical Science (ISAS), Japan Aerospace Exploration Agency (JAXA), Sagamihara, Kanagawa 252-0222, Japan}
\altaffiltext{16}{International School for Advanced Studies (SISSA), Via Bonomea 265, 34136, Trieste, Italy}
\altaffiltext{17}{Kavli Institute for Cosmology Cambridge, Cambridge CB3 OHA, UK}
\altaffiltext{18}{Kavli Institute for the Physics and Mathematics of the Universe (Kavli IPMU, WPI), UTIAS, The University of Tokyo, Kashiwa, Chiba 277-8583, Japan}
\altaffiltext{19}{Kavli Institute for the Physics and Mathematics of the Universe (WPI), Berkeley Satellite, the University of California, Berkeley 94720, USA}
\altaffiltext{20}{National Institute for Nuclear Physics (INFN), Sezione di Trieste Via Valerio 2, I-34127, Trieste, Italy}
\altaffiltext{21}{Physics Division, Lawrence Berkeley National Laboratory, Berkeley, CA 94720, USA}
\altaffiltext{22}{Radio Astronomy Laboratory, University of California, Berkeley, CA 94720, USA}
\altaffiltext{23}{Research Center for the Early Universe, School of Science, The University of Tokyo, Tokyo 113-0033, Japan}
\altaffiltext{24}{SOKENDAI (The Graduate University for Advanced Studies), Shonan Village, Hayama, Kanagawa 240-0193, Japan}
\altaffiltext{25}{School of Physics, University of Melbourne, Parkville, VIC 3010, Australia}
\altaffiltext{26}{Space Sciences Laboratory, University of California, Berkeley, CA 94720, USA}
\altaffiltext{27}{Yokohama National University, Yokohama, Kanagawa 240-8501, Japan}

\begin{abstract}

We present a measurement of the gravitational lensing deflection power spectrum reconstructed with two seasons cosmic microwave background polarization data from the \pb\ experiment.
Observations were taken at 150 GHz from 2012 to 2014 which survey three patches of sky totaling 30 square degrees.
We test the consistency of the lensing spectrum with a Cold Dark Matter (CDM) cosmology and reject the no-lensing hypothesis at a confidence of \nolenssig\ including statistical and systematic uncertainties.
We observe a value of \Alens\ = \alensvalanderr\ using all polarization lensing estimators, which corresponds to a 24\% accurate measurement of the lensing amplitude.
Compared to the analysis of the first year data, we have improved the breadth of both the suite of null tests and the error terms included in the estimation of systematic contamination.

\end{abstract}

\keywords{}

\section{Introduction}

The polarization of the cosmic microwave background (CMB) not only gives us insight into the earliest stages in the evolution of the universe, it also allows us to probe the large scale structure (LSS) formed more recently in cosmological history. CMB polarization can be separated into even parity $E$-modes and odd parity $B$-modes, and while the $E$-modes can be sourced from the same scalar perturbations that dominate CMB temperature anisotropies, $B$-modes are not generated through this mechanism to first order in perturbations.

Much effort is being devoted to using CMB $B$-modes for signs of primordial gravitational waves, but another expected source of $B$-modes is the gravitational lensing of the CMB by LSS \cite[]{2006PhR...429....1L}.
This signature appears in the $B$-mode power spectrum as a signal peaking at an angular scale $\ell\sim 1000$. By mapping the CMB polarization, we can extract information about the distribution of LSS through reconstruction of the CMB lensing potential.

The CMB lensing potential is a representation of the matter power spectrum, integrated along the line of sight of CMB photons, which can tell us how much a given photon will be gravitationally deflected. For a gravitational potential $\Psi$ we can integrate along the line of sight to calculate the lensing potential $\phi$ \cite[]{HO},
\beq \phi(\bm n) = -2\int_0^{\chi_*} d\chi \frac{\chi_* - \chi}{\chi_* \chi} \Psi(\chi \bm n, \chi), \eeq
where $\chi$ is the comoving distance and $\chi_*$ is the comoving distance to the surface of last scattering.
The lensing potential is related to the deflection field $\bm d = \bm \nabla \phi$, which tells us how much a photon of the CMB is gravitationally deflected across the sky as it travels from the surface of last scattering to our detector.

We are able to reconstruct the lensing potential by taking advantage of the statistical properties of the CMB. At the surface of last scattering, the CMB is well described as a statistically isotropic Gaussian random field, but gravitational lensing introduces non-Gaussianities that correlate CMB modes of different angular scale. This non-Gaussianity allows us to reconstruct the underlying lensing potential $\phi$ by correlating $E$- and $B$-modes at varying angular scale \cite[]{HO}.

The science of CMB lensing contains a wealth of information about the more recent evolution of the universe, including the formation of LSS and the physics of neutrinos \cite[]{2009AIPC.1141..121S}.
The polarized CMB in particular is promising as a tracer of LSS because $B$-modes are not dominated by cosmic variance of the primordial CMB in the same way that the temperature and $E$-modes are at the present time. Additionally, polarization measurements are also less affected by many of the sources of contamination for the CMB temperature anisotropies, e.g. from the atmosphere or extragalactic foregrounds like the CIB and SZ-effects \cite[]{osborne2014}.

The lensing potential has been detected using both CMB temperature and polarization fluctuations by a number of experiments including \pb\ -- from the first season dataset \cite[]{pb2014b}, \bicep2/\keckarray\ \cite[]{Array:2016afx}, \actpol\ \cite[]{Sherwin:2016tyf}, Planck \cite[]{2018arXiv180706210P}, and \sptpol\ \cite[]{Wu:2019hek}.

%Additionally, cross correlations between the CMB lensing potential with external tracers has been demonstrated in other \pb\ analyses including cross correlation with the cosmic shear measurement of the \hsc\ \cite[]{Namikawa_2019} and the sub-mm galaxy measurements of the \herschel\ \cite[]{Faundez:2019lui} experiments. These cross correlations are valuable in that they can combine information from two independent tracers of LSS while avoiding instrument-specific systematic errors \cite[]{Bianchini_2015}.

\edit{Additionally, cross correlations between the CMB lensing potential with external tracers have been carried out in other works. These are valuable for combining information from two independent tracers of LSS while avoiding instrument-specific systematic errors \cite[]{Bianchini_2015}.

Cross correlation with cosmic shear has been demonstrated by a number of experiments \cite[]{PhysRevD.91.062001, 10.1093/mnras/stw570, PhysRevD.93.103508, 10.1093/mnras/stx1828, 2016MNRAS.460.1270D, 2017MNRAS.465.1454H, PhysRevD.100.043517}, deriving results primarily from \cmb\ temperature. The dataset of this paper has also been used in a cosmic shear cross correlation with \hsc\ \cite[]{Namikawa_2019}. Cross correlation with the cosmic infrared background have been conducted as well \cite[]{Holder_2013, Planck_CIB_cross, pb2014a, Engelen_2015, 2018arXiv180706210P}. Additionally, this dataset has been cross correlated with sub-mm galaxy counts from the \herschel\ \cite[]{Faundez:2019lui} experiment.}

The search for CMB $B$-modes from gravitational waves can be improved if the $B$-mode signal from gravitational lensing is reduced.
This ``delensing'' has been done using several methods.
External tracers of the lensing potential have been combined with CMB observations \cite[]{Manzotti_2017, PhysRevD.92.043005} to subtract templates of gravitational lensing and reduce the final $B$-mode power. Internal delensing has also been achieved in which the lensing potential and $B$-modes are constructed using the same dataset \cite[]{Carron_2017}, and in another work we demonstrate internal delensing of the CMB using only polarization data \cite[]{2019arXiv190913832A}. Both of these delensing methods are useful, but of the two, internal delensing has been forecast to achieve the best performance for sufficiently low noise measurements \cite[]{PhysRevD.99.043518}.

In this work we show a reconstruction of the lensing potential power spectrum from observations by the \pb\ experiment.
We have observed an area of $\sim 30$ square degrees with one of the lowest levels of arc-minute scale noise yet achieved.
The lensing information is dominated by polarization rather than temperature anistotropies. This deep data set has enabled a polarization-only reconstruction of the lensing potential power spectrum, and has served as a useful dataset for additional cross correlation and delensing studies.

\section{Lensing Power Spectrum Analysis} \label{sec:analysis}

The polarization-sensitive \pb\ experiment is located at the James Ax Observatory in Northern Chile on Cerro Toco. It uses 1,274 transition-edge sensor bolometers to observe the CMB at 150 GHz and has a 2.5 meter primary mirror that produces a beam with a 3.5 arcmin full width at half maximum (FWHM).

We observe three sky patches over a time period of two years from 2012 to 2014, each with an extent of approximately $3^\circ \times 3^\circ$. They are centered in right ascension and declination at ($4^\mathrm{h}40^\mathrm{m}12^\mathrm{s}, -45^\circ 00'$), ($11^\mathrm{h}53^\mathrm{m}0^\mathrm{s}, -0^\circ 30'$) and ($23^\mathrm{h}1^\mathrm{m}48^\mathrm{s}, -32^\circ 48'$) which we will refer to with the respective names RA4.5, RA12 and RA23. More details on the receiver and telescope can be found in \cite[]{Arnold_SPIE2012, Kermish_SPIE2012}. One advantage of observing small patches is the ability to obtain deeper observations over a given amount of time.
%The effective white noise levels after accounting for beam and filter transfer functions are 10 $\mu$K-arcmin for RA4.5, 7 $\mu$K-arcmin for RA12 and 6 $\mu$K-arcmin for RA23.
\edit{The polarization white noise levels for RA4.5, RA12 and RA23 respectively are 7 $\mu$K-arcmin, 6 $\mu$K-arcmin and 5 $\mu$K-arcmin.
After accounting for beam and filter transfer functions, the effective polarization noise levels (defined as the minima of the resulting $N_\ell$ noise curves) are 10 $\mu$K-arcmin, 7 $\mu$K-arcmin and 6 $\mu$K-arcmin, respectively.}

This analysis builds on previous results from the \pb\ collaboration using the same dataset described above. We have shown evidence of $B$-mode power induced by gravitational lensing \cite[]{pb2017a}, which we will refer to as \spp. The CMB maps used in that analysis are also used here.

We also showed evidence of the lensing potential auto-power spectrum itself in a previous work \cite[]{pb2014b} that used only our first season of observations. We will refer to that paper as \ipplensing. This paper improves upon that work by adding a second year of observations on the same set of three patches, which corresponds to an increase in data volume of 61\% over \ipplensing.

We also note that in \ipplensing, we used a separate analysis pipeline from our $B$-mode analysis to generate simulations and perform null tests. This time our analysis uses the same pipeline to generate lensed and filtered CMB simulations as used in \spp, so that the details of our mapmaking and instrumental systematic estimation are consistent across both publications. This has the advantage that our simulations now accurately model our mapmaking procedure starting at the timestream level and include the anisotropic effects of our timestream filters in the lensing reconstruction step. Additionally we have included a set of data split null tests not present in our first season lensing analysis, these are described in more detail in Section~\ref{sec:split}.

In our data analysis pipeline, we start with $Q$ and $U$ CMB maps to obtain weighted $E$- and $B$-modes using the data model
\beq d_i = P_{ik}s_k + n_i, \label{eq:data_model} \eeq
where $d_i$ contains the pixelized real space $Q$ and $U$ maps, $n_i$ are the pixelized map domain noise contributions, and $s_k$ are the $E$- and $B$-mode fields. $P_{ik}$ is the matrix operator that encodes effects from the beam and timestream filtering, and transforms from Fourier space to real space. The index $i$ includes $Q/U$ and pixel indices $i=(M,p)$, and the index $k$ includes $E/B$ and mode indices $k=(X,\Ell)$.

We obtain inverse-variance Wiener-filtered CMB $E$- and $B$-modes, $\overline X(\Ell)$, from the observed $Q$ and $U$ maps, $d$, using the matrix equation
\beq \overline X = S^{-1} \left[ S^{-1} + P^\dagger N^{-1} P \right]^{-1} P^\dagger N^{-1} d, \eeq
where $S_{kk'} = \delta_{XX'}\delta_{\Ell\Ell'}C_\ell^{XX}$ and $N_{ii'} = \delta_{MM'}\delta_{pp'}N_p^M$. $C_\ell^{XX}$ are the fiducial CMB power spectra for $X\in\{E,B\}$ and $N_p^M$ is the noise map where $p$ labels a given pixel in the map and $M\in\{Q,U\}$. Our noise weighting also includes a cutoff for pixels with noise levels above 55 $\mu$K-arcmin and point source masking for sources above 25 mJy in intensity. The CMB power spectra used for this Wiener filter are generated using the freely available software package CAMB\renewcommand{\thefootnote}{$\dagger$}\footnote{\href{https://camb.info/}{camb.info}}, and use the Planck 2015 best fit cosmological parameters \cite[]{2015PlanckXIII}\renewcommand{\thefootnote}{$\ddagger$}\footnote{in the \texttt{base\_plikHM\_TT\_lowTEB\_lensing} configuration}, which is the same parameter set used in \spp.\renewcommand{\thefootnote}{\arabic{footnote}}

From the inverse variance weighted modes $\overline X(\Ell)$ we then reconstruct the lensing potential using the quadratic estimator
\beq \hat\phi^{XY}(\LL) = A(\LL)\int d^2\Ell\ \overline X(\Ell)\ \overline Y^*(\Ell-\LL)\ F_{XY}(\Ell, \Ell-\LL), \eeq
where the normalization is defined by
\beq A^{-1}(\LL) = \int d^2\Ell\ f_{XY}(\Ell, \Ell-\LL) F_{XY}(\Ell, \Ell-\LL), \eeq
and the weights $f_{XY}(\Ell, \Ell-\LL)$ and $F_{XY}(\Ell, \Ell-\LL)$ are described in detail in \cite[]{Hu_2002}.

In addition to our data we also use a set of 500 Monte Carlo (\mc) simulations in our analysis to estimate the lensing mean field, noise bias, transfer function and covariance matrix. We generate realizations of lensed CMB signal that are mock observed using the same pointing, noise level and scan strategy as our real observations. These timestreams are then run through our mapmaking pipeline and the resulting $Q$ and $U$ maps are used as inputs to our lensing pipeline as described in the above equations.

The process of going from quadratic estimates of the lensing potential $\hat\phi^{XY}(\LL)$ to power spectra follows the method we used in \ipplensing. First we estimate the mean field from our set of \mc\ simulations and subtract that from $\hat\phi^{XY}(\LL)$. Next we correlate two reconstructed lensing maps $\hat\phi^{UV}(\LL)$ and ${\hat\phi}^{XY}(\LL)$ to construct the pseudospectra
$\hat C_L^{UVXY}$,
where the indices $UV,XY$ indicate the type of estimator ($EE$ or $EB$).
We follow the \cite[]{hanson2011} and \cite[]{Namikawa2013} to estimate the realization-dependent noise bias $N_L^{(0),UVXY}$ using lensing reconstructions of our data and \mc\ simulations.
Once bias subtracted spectra from simulations are constructed, we then estimate the transfer function by taking the ratio between the mean of these reconstructed lensing power spectra and the input theory power spectrum used to generate them. And finally, this transfer function is used to correct the lensing potential power spectrum estimate of our data giving us our final spectra as defined by the equation,
\beq C_L^{UVXY} = \left.\left(\hat C_L^{UVXY} - N_L^{(0),UVXY}\right)\right/T_L. \label{eq:transfer_and_bias} \eeq
Here, $T_L$ is the transfer function that corrects for the effects of filtering and weighting in our pipeline and $C_L$ is the lensing potential power spectrum. The lensing estimators are labeled here by $UV, XY \in \{EE, EB\}$. Additionally, while we only used $C_L^{EEEB}$ and $C_L^{EBEB}$ in \ipplensing, we include the power spectrum estimator $C_L^{EEEE}$ in this analysis.

\edit{We also considered including an estimate of the $N^{(1)}$ bias but ultimately did not use it for this analysis because its expected size is small relative to our lensing spectrum sensitivity. We compared analytical estimates of this bias to the size of our statistical errors for each of the lensing estimators and found that the relative size of the $N^{(1)}$ bias is only a few percent. Estimates of this bias were calculated with publicly available software\renewcommand{\thefootnote}{$*$}\footnote{\href{https://github.com/JulienPeloton/lensingbiases}{github.com/JulienPeloton/lensingbiases}}\renewcommand{\thefootnote}{\arabic{footnote}} using numerical methods shared by other codes \cite[]{PhysRevD.96.063510}.}

To estimate the amplitude of lensing, we use 500 \mc\ simulations to construct the covariance between our three $C_L^{UVXY}$ estimators. If we label the estimator $\alpha=UVXY\in\{EEEE,EEEB,EBEB\}$, and the covariance matrix $\cov$ represents the covariance between the bandpower $C_L^\alpha$ and $C_{L'}^{\alpha'}$, then the lensing amplitude is
\beq \Alense = \frac{\sum_{L\alpha L'\alpha'} C_L^{\alpha\mathrm{(th)}} \invcov C_{L'}^{\alpha'}}{\sum_{L\alpha L'\alpha'} C_L^{\alpha\mathrm{(th)}} \invcov C_{L'}^{\alpha'\mathrm{(th)}}}, \eeq
and the inverse variance on the amplitude is given by
\beq (\sigma_A)^{-2} = \sum_{L\alpha L'\alpha'} C_L^{\alpha\mathrm{(th)}} \invcov C_{L'}^{\alpha'\mathrm{(th)}}, \eeq
where the (th) \edit{superscript} denotes the theory power spectrum.

Finally we have also found that our observations are polarization dominated. While we do not include temperature in the results presented here, we have compared $N^{(0)}$ bias curves from temperature-only information (the $TT$ estimator) and from polarization-only information (the $EE$ and $EB$ estimators) and found lower noise in polarization.

\section{Null Tests}

We perform a blind analysis and therefore need a way to guard against unknown systematics and validate our error-bar estimation, which we address through a set of null tests.
We only examined our final power spectra after all of our null tests satisfy passing criteria that demonstrate our analysis is performing as expected.

All of the simulations used in the following null tests are generated at the timestream level. They use the same pointing reconstruction used for real observations to mock simulate CMB signal sky observations, and include noise at the timestream level based on a white noise model consistent with the real observations of the second season in \spp. The resulting simulated timestreams are run through the same mapmaking and lensing reconstruction pipelines as is used for the real data.

\subsection{Data Split Null Tests} \label{sec:split}

We perform one suite of null tests constructed from splits in our data selection. We choose twelve splits to probe potential systematic errors that are not captured by the lensing analysis pipeline.
The splits are listed in Table~\ref{tab:splits}.
These are the same data splits used in \spp, where more detailed description of the twelve data splits can be found.

\begin{table}[htbp]
\begin{center}
\caption{\label{tab:splits} Data split null test types.}
\begin{tabular}{c}
\hline
First vs. second seasons of data collection \\
Close to sun vs. far from sun \\
Day vs. night \\
First half vs. second half by data volume \\
Rising vs. setting \\
High elevation vs. low elevation \\
High vs. low detector gain \\
Good vs. bad weather \\
Q vs. U pixels \\
Left vs. right side of the focal plane \\
Left vs. right scan direction \\
Close to moon vs. far from moon \\
\hline
\end{tabular}
\end{center}
%\tablecomments{
The twelve ways that we split the dataset for null tests that probe potential unmodeled systematic errors.
%}
\end{table}

For each of these twelve data splits we construct two sets of lensing estimates, $\hat\phi^{U_1V_1}(\LL)$ and $\hat\phi^{X_1Y_1}(\LL)$ for the first set of the split dataset and $\hat\phi^{U_2V_2}(\LL)$ and $\hat\phi^{X_2Y_2}(\LL)$ for the second set of the split dataset. We then construct the auto spectra of each of the two sets and the cross spectra between the two sets, including a noise bias subtraction for each of these component spectra. Finally we use these to construct the null spectrum
\beq C_L^{\mathrm{null}} = C_L^{U_1V_1X_1Y_1} + C_L^{U_2V_2X_2Y_2} - C_L^{U_1V_1X_2Y_2} - C_L^{U_2V_2X_1Y_1}. \eeq

We evaluate this set of 108 null spectra (from 12 splits, 3 power spectrum estimators, and 3 sky patches) similarly to the procedure used in \spp.

Using nine equally spaced bins $b$ in the multipole range $100 < L < 1900$ and an estimate of the standard deviation $\sigma_b$ from \mc\ simulations we construct the quantity \chinullb{} $\equiv C_b^{\mathrm{null}}/\sigma_b$. For each patch, we then calculate the probability to exceed (PTE) value for five quantities: the average value of \chinullb, the worst value of \chisqnullb, the worst value of \chisqnull\ by spectrum (summed over all bins), the worst value of \chisqnull\ by test, and finally the total value of \chisqnull\ for each patch. The simulated data which are generated from the simulated timestreams are split the same way as the observed data. The error bars $\sigma_b$ are then estimated from an ensemble of 500 simulated data splits. The results from these null tests are summarized in Table~\ref{tab:pte_summary_split}.

%Our unblinding criteria involves calculating a total PTE (labeled as ``All stats'' in Table~\ref{tab:pte_summary_split}) for the five tests just described.
%We take the worst of the five $\chi^2$ PTEs from the data, and then we compare to the worst PTE from a distribution of simulations. The ``All stats'' PTE is the fraction of the simulations that exceed the data and we require that this total PTE is above 5\% before ever calculating the final power spectra from data.
%The results from these null tests are summarized in the rightmost column of Table~\ref{tab:pte_summary_split}.

%For the PTEs involving $\chi^2$ statistics we calculate $\chi^2_\mathrm{data}$ for our data and $\chi^2_{\mathrm{sim},i}$ for each of our 500 \mc\ simulations.
%Our PTE value is then equal to the fraction of the simulations such that $\chi^2_{\mathrm{sim},i} > \chi^2_\mathrm{data}$.
%For the average \chinullb\ statistic we evaluate the PTE slightly differently than the rest by performing a two sided test.
%We calculate the average $\chi_\mathrm{data}$ for the data and the average $\chi_{\mathrm{sim},i}$ for each simulation and the corresponding PTE is equal to the fraction of the simulations such that $\chi_{\mathrm{sim},i}>\chi_\mathrm{data}$.

\edit{Before unblinding our data we summarize the five tests just described by calculating a total PTE, labeled ``All stats'' in the rightmost column of Table~\ref{tab:pte_summary_split}. We require this value to be greater than 5\%.
To calculate this, we take the worst of the five $\chi^2$ PTEs from the data (in each row of Table~\ref{tab:pte_summary_split}) and compare it to the worst PTE from a distribution of simulations. The resulting ``All stats'' PTE is then the fraction of the simulations that exceed the data.}

\edit{The worst $\chi^2$ PTEs are calculated in a similar manner. We calculate one value of $\chi^2_\mathrm{data}$ from the data and 500 values of $\chi^2_{\mathrm{sim},i}$ from a distribution of \mc\ simulations.
The PTE value is then equal to the fraction of the simulations such that $\chi^2_{\mathrm{sim},i} > \chi^2_\mathrm{data}$.
The only exception to this rule is the average \chinullb\ PTE, which we evaluate by performing a two sided test.
We calculate the average $\chi_\mathrm{data}$ from data and the average $\chi_{\mathrm{sim},i}$ for each simulation and the corresponding PTE is equal to the fraction of the simulations such that $\chi_{\mathrm{sim},i}>\chi_\mathrm{data}$.}

\begin{table*}[htbp]
\begin{center}
\caption{\label{tab:pte_summary_split}PTEs resulting from the data split null tests.}
\begin{tabular}{c|ccccc|c}
\hline
\hline
 Patch & \shortstack{Average of \\ \chinullb}
       & \shortstack{Extreme of \\ \chisqnullb}
       & \shortstack{Extreme of \\ \chisqnull\ by spectrum}
       & \shortstack{Extreme of \\ \chisqnull\ by test}
       & \shortstack{Total      \\ \chisqnull}
       & \shortstack{All        \\ stats} \\
\hline
 RA4.5 & 47.8\% & 58.8\% & 56.2\% & 92.6\% & 99.0\% & 75.0\% \\
 RA12  & 47.2\% & 43.8\% & 85.2\% & 76.8\% & 50.0\% & 73.0\% \\
 RA23  & 35.8\% & 78.2\% & 81.4\% & 69.6\% & 47.4\% & 61.2\% \\
\hline
\hline
\end{tabular}
\end{center}
%\tablecomments{
PTEs resulting from the data split null tests. The furthest right column capturing the results of all five tests are above 5\% indicating that our dataset passes the null test criteria. We also checked to see if the distribution of all PTEs agree with a uniform distribution via the Kolmogorov-Smirnov test and found that for all of our patches the null PTEs are indeed consistent with a uniform distribution as expected. One notable feature of the table is a preponderance of high PTE values which is caused by our treatment of the noise bias subtraction. We elaborate on this in the appendix but otherwise we find no evidence for systematic biases in the data.
%\ref{app:split_ptes}.
%}
\end{table*}

\subsection{Curl and Cross-Patch Null Tests}

Additionally we conduct a set of lensing specific null tests using the full dataset. First we generate curl reconstructions of the lensing deflection field ${\bm\nabla}\times{\bm d}(\hat{\bm n})$, which we expect to be vanishingly small and serve as a check on unmodeled systematics \cite[]{cooray2005curl}.

We also generate cross power spectra between lensing reconstructions from two different observational patches. These independent measurements should lack any common signal, so any significant deviation from a null spectrum would indicate a misestimation of our error bars or a spurious correlation introduced by our analysis pipeline.

Both of these tests were also performed in \ipplensing. Our passing criteria for these sets of tests are similar to our criteria for the data split null tests. We calculate the worst $\chi^2$ PTEs corresponding to the average of \chinullb, extreme of \chisqnullb, extreme of \chisqnull\ by spectrum, and a total \chisqnull, in addition to a combined PTE combining all four of those statistics. We consider the dataset to have passed these tests if the final PTE accounting for all statistics is greater than 5\%. The results from this set of null tests are summarized in Table~\ref{tab:pte_summary_curl_cross}, in particular showing that the PTEs for all statistics are 53.0\% for the curl tests and 60.2\% for the cross-patch tests.
We also note that these null tests do not require a noise bias subtraction and thus are not affected by the noise bias calculation subtlety described in the Appendix.

\begin{table*}[htbp]
\begin{center}
\caption{\label{tab:pte_summary_curl_cross}PTEs resulting from the curl and cross-patch null tests.}
\begin{tabular}{c|cccc|c}
\hline
\hline
 Test  & \shortstack{Average of \\ \chinullb}
       & \shortstack{Extreme of \\ \chisqnullb}
       & \shortstack{Extreme of \\ \chisqnull by spectrum}
       & \shortstack{Total      \\ \chisqnull}
       & \shortstack{All        \\ stats} \\
\hline
 curl  & 48.6\% & 23.8\% & 58.6\% & 64.8\% & 53.0\% \\
 cross & 95.4\% & 16.0\% & 27.4\% & 16.2\% & 60.2\% \\
\hline
\hline
\end{tabular}
\end{center}
%\tablecomments{
PTEs resulting from the curl and cross-patch null tests. Again we see that all of the individual worst \chisqnull\ criteria and the PTE combining all stats in the rightmost column are all above the required null test threshold. Like the data split tests, we also tested the distribution of PTEs and found they are consistent with a uniform distribution via the Kolmogorov-Smirnov test.
%}
\end{table*}

\section{Contamination}

We use a difference spectrum framework to evaluate the effect of instrumental systematic and foreground contamination to the lensing spectrum by looking at the effect on \Alens.

Using a set of \mc\ simulations, we calculate two lensing power spectra for each CMB realization. The first spectrum is created with the fiducial pipeline while the second spectrum is created by
adding a realization of the contamination at map level to our $Q$ and $U$ maps.
The difference between these two is used as our estimate of contamination,
\beq \Delta C_L^c = C_L^c - C_L, \eeq
where $C_L^c$ denotes the lensing power spectrum calculated including contamination while $C_L$ is the spectrum calculated without contamination.

\subsection{Instrumental Systematics}

In \spp\ we used simulations of systematic effects to estimate their contributions to the \clbb\ power spectrum. This systematics pipeline was incorporated into our main analysis pipeline and generated contamination at the timestream level that modeled a number of different instrumental systematic effects. We use that same systematic simulation pipeline here to estimate contributions to the lensing power spectrum.

For each instrumental systematic effect we use 100 \mc\ estimates of $\Delta C_L^c$. The mean value of these spectra and their covariance are then used to calculate an effective lensing amplitude due to systematic contamination, \Alenscont{} $\pm$ \sigmaAlenscont{}. To evaluate any bias introduced to \Alens\ by a given systematic effect, we calculate an upper limit \deltaAlenscont\ on the lensing amplitude given by
\beq \Delta \Alensconte = |\Alensconte| + \frac{\sigmaAlensconte}{\sqrt{100}} . \eeq

In addition to limits on systematic bias to the lensing power spectrum, we also account for the extra variance introduced by systematic effects through their effects on \sigmaAlens. We add in quadrature all the values of \sigmaAlenscont\ in our final estimation of \Alens.

A summary of the contributions \sigmaAlenscont\ and \deltaAlenscont\ from each systematic effect is shown in Table~\ref{tab:systematics}, in particular showing that the total contribution to \sigmaAlenscont\ is 0.02 and our upper limit on systematic bias from all modeled effects is 0.006.

\begin{table}[htbp]
\begin{center}
\caption{\label{tab:systematics} Contributions to \Alens\ from instrumental systematic effects.}
\begin{tabular}{lrrr}
\hline
Effect \edit{[$\times 10^{3}$]} & \Alenscont & \sigmaAlenscont & \deltaAlenscont \\
\hline
Crosstalk           & -0.28 & 1.9   & 0.47 \\
Pointing            & 3.60  & 21.2  & 5.72 \\
Beam Ellipticity    & 0.54  & 1.5   & 0.69 \\
Beam Size           & 0.16  & 1.8   & 0.34 \\
Gain Drift          & 0.14  & 2.7   & 0.41 \\
Relative Gain       & -0.67 & 4.9   & 1.16 \\
\hline
Total               &       & 22.1  & 5.92 \\
\hline
\end{tabular}
\end{center}
%\tablecomments{
All values have been multiplied by a factor of $10^3$ for display in this table. The resulting total contribution to our uncertainty on the lensing amplitude is \sigmaAlenscont{} $=0.022$, and our upper limit on the total systematic bias is \deltaAlenscont{} $=0.006$.
%}
\end{table}

\subsection{Foregrounds}

We use the Planck 2015 frequency maps to estimate the impact that foregrounds have on our reconstruction of gravitational lensing \cite[]{2016A&A...594A...9P, 2016A&A...594A..10P}. In particular, we use the Planck 30 GHz and 353 GHz all sky intensity maps as tracers of synchrotron and dust foreground power, respectively. Our observational patches were chosen, in part, because they have very low foreground power. Thus, the Planck polarization maps are dominated by noise in the regions of the sky that we observed. Therefore, to estimate a conservative upper limit on foreground power, we use a polarization fraction of $p=20\%$ and constant polarization angle in combination with Planck intensity maps to generate maps of polarized foregrounds in our three patches.

\edit{This method does have certain drawbacks. The synchrotron estimate in particular has limited information of smaller scales due to the half degree beamwidth of the Planck 30 GHz observations. However, selecting for regions of very low foreground power means that an estimate from a method like simulations from a fitted foreground template would not be representative of these regions so we instead use direct observations as a tracer. Similarly, a constant polarization angle is only an approximation but since polarization angles are coherent across scales larger than our patch we use a constant angle as a close approximation in the absence of high quality measurements of the underlying angle distributions.

Additionally, the overall estimates of dust and synchrotron power in these patches are respectively three and five orders of magnitude smaller than the $E$-mode power which dominates the reconstructions and at least two orders of magnitude smaller than the noise power. Even if there were moderate errors in foreground estimation, we can sensibly say that the overall contributions to the lensing spectra would be negligible. And as a final check, the curl null tests previously described are sensitive to foreground power and the passing of these tests indicates that there is not a significant level of foreground contamination.}

The amplitudes of these Planck maps are scaled to 150 GHz assuming a modified blackbody spectral dependence for thermal dust and a power law for synchrotron \cite[]{2018A&A...618A.166K, 2018arXiv180104945P}, and then simulated timestreams are produced and run through our analysis pipeline in order to include the scan strategy, time stream processing, filtering, and other effects that are incorporated in our real observations.

Finally, contributions to the lensing power spectrum $\Delta C_L^c$ for our dust and synchrotron estimates are constructed using the same method as the instrumental systematics, and contributions to the bias and uncertainty on \Alens\ are calculated and listed in Table~\ref{tab:foregrounds}.

\begin{table}[htbp]
\begin{center}
\caption{\label{tab:foregrounds} Contributions to \Alens\ from foreground contamination.}
\begin{tabular}{lrrr}
\hline
Effect \edit{[$\times 10^{3}$]} & \Alenscont & \sigmaAlenscont & \deltaAlenscont \\
\hline
Dust                   &  3.16 & 65.1  & 9.67 \\
Synchrotron            & -0.42 &  7.6  & 1.18 \\
\hline
Total                  &       & 65.5  & 9.74 \\
\hline
\end{tabular}
\end{center}
%\tablecomments{
All values have been multiplied by a factor of $10^3$ for display in this table. The resulting total contribution to our uncertainty on the lensing amplitude is \sigmaAlenscont{} $=0.066$, and our upper limit on the total systematic bias is \deltaAlenscont{} $=0.0097$.
%}
\end{table}

\section{Results}

\begin{figure}
\begin{center}

\includegraphics[width=0.5\textwidth]{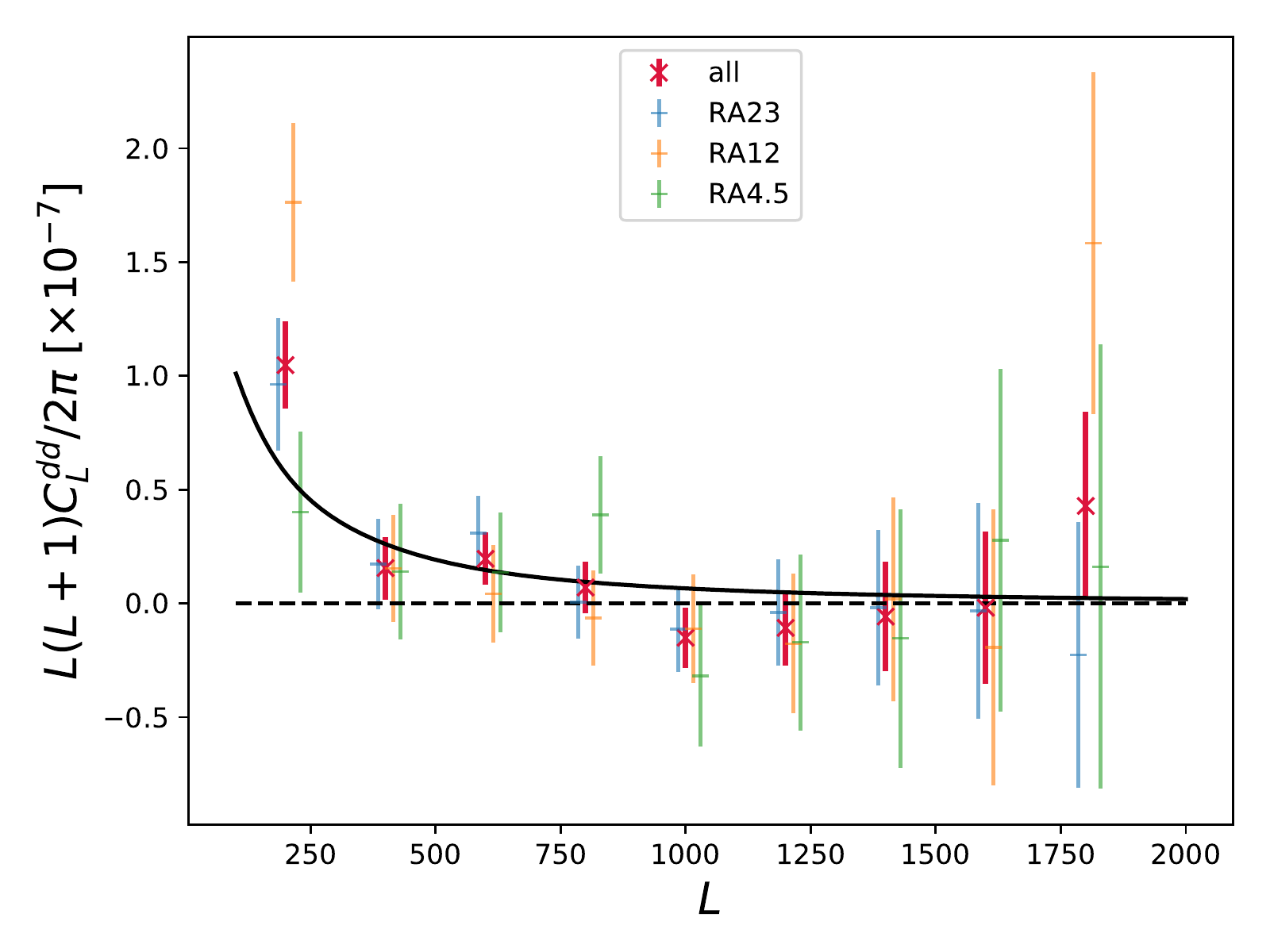}

\caption{\label{fig:spectrum} Minimum variance lensing deflection power spectrum, with variance taken from the diagonal elements of the covariance matrix. The black solid curve represents the power spectrum for \Alens$=1$. Red data points are the minimum variance \pb\ power spectrum from a combination of our three observational patches and three power spectrum estimators, $C_L^{EEEE}$, $C_L^{EEEB}$ and $C_L^{EBEB}$. The blue, orange, and green points represent the power spectra for each of the three patches (RA23, RA12, and RA4.5 respectively) and are offset in $L$ in the above plot for clarity.}

\end{center}
\end{figure}

We present the minimum variance power spectrum in Figure~\ref{fig:spectrum}, which combines power spectra from our three observational patches and the three polarized estimators. The bandpowers and error bars are listed in Table~\ref{tab:mv_spectrum}. The statistical uncertainty on our measurement of \Alens\ is calculated from the standard deviation of the distribution of simulated \Alens\ from 500 signal-plus-noise \mc\ simulations. Including uncertainty from instrumental systematics and foreground contamination, our measurement of the lensing amplitude is \alensvalanderr, corresponding to a significance of \lenssig.

\begin{table}[htbp]
\begin{center}
\caption{\label{tab:mv_spectrum} Minimum variance spectrum bandpowers.}
\begin{tabular}{r|r}
\hline
Central $L$ & $D_L[\times 10^{-8}]$ \\
\hline
200  & $10.47 \pm 1.91$ \\
400  & $ 1.55 \pm 1.38$ \\
600  & $ 1.96 \pm 1.15$ \\
800  & $ 0.70 \pm 1.14$ \\
1000 & $-1.51 \pm 1.33$ \\
1200 & $-1.08 \pm 1.67$ \\
1400 & $-0.58 \pm 2.41$ \\
1600 & $-0.19 \pm 3.35$ \\
1800 & $ 4.28 \pm 4.13$ \\
\hline
\end{tabular}
\end{center}
%\tablecomments{
The minimum variance power spectrum $D_L = L(L+1)C_L/2\pi$ with $1\sigma$ error bars, multiplied by a factor of $10^8$ in this table for display purposes.
%}
\end{table}

\begin{figure*}
\begin{center}

\includegraphics[width=\textwidth]{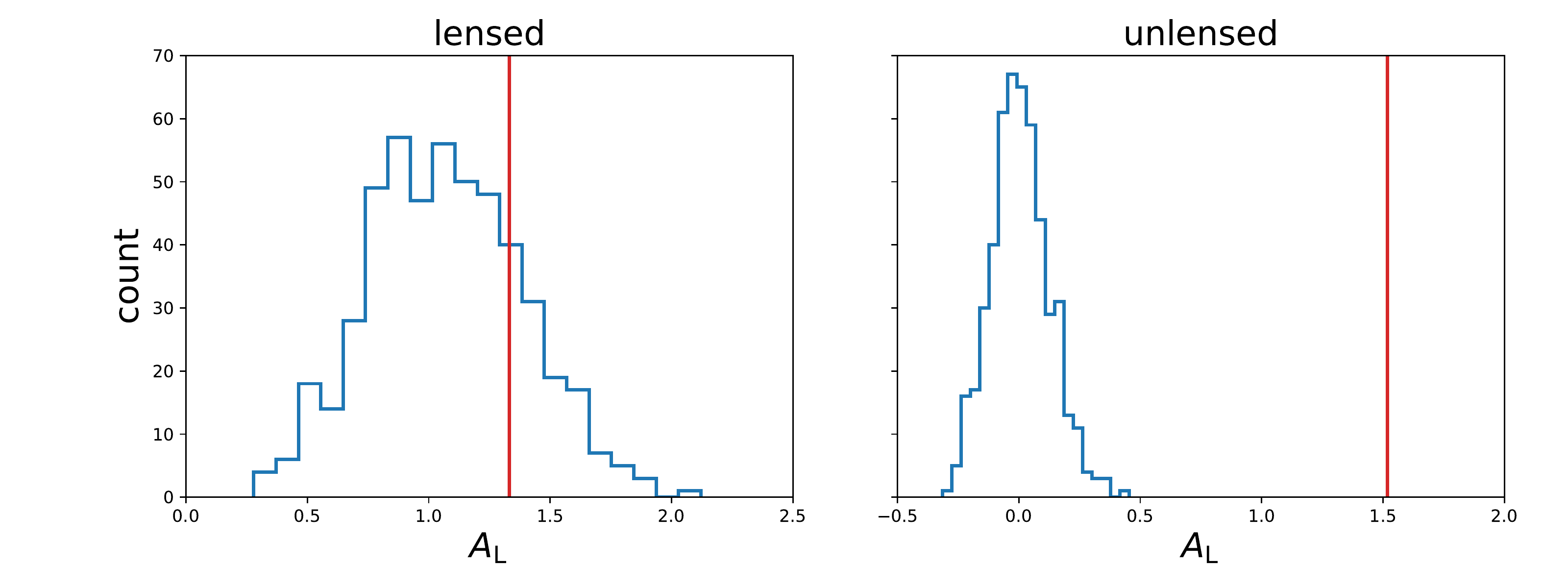}

\caption{\label{fig:histograms} Distribution of \Alens\ from 500 \mc\ simulations compared to the observed amplitude. The right plot assumes no lensing while the left plot uses lensed CMB simulations. In both cases, the blue histogram represents the distribution of \Alens\ and the red vertical line marks our observed value. As in \ipplensing, different weighting in the lensing estimator under the lensing/no-lensing cases results in different values of \Alens.}

\end{center}
\end{figure*}

Additionally, we examine the no-lensing hypothesis using a set of 500 \mc\ simulations that do not include gravitational lensing.
The distribution of unlensed simulations has a width of \sigmaAlens{} $ = 0.12$, corresponding to a forecasted significance of $8.3\sigma$.
%As was the case in \ipplensing, the suboptimal weighting in the lensing estimator due to the assumption of no-lensing has the effect of shifting the value of the lensing amplitude.
\edit{The suboptimal weighting in the lensing estimator due to the assumption of no-lensing has the effect of shifting the value of the lensing amplitude.}
The power spectrum calculated under the no-lensing assumption on our data has an estimated amplitude of \Alens{} $ = 1.52$.
\edit{The shift in \Alens\ from the lensed to unlensed case here is similar to the shift seen between the two \Alens\ values reported in \ipplensing.}
Including uncertainty from systematics and foregrounds and our observed value of \Alens, we reject the no-lensing hypothesis at a significance of \nolenssig, which is a considerable improvement upon the $4.2\sigma$ rejection from our earlier work in \ipplensing.
Distributions of the \Alens\ calculated from simulations in the lensed and unlensed cases are shown in Figure~\ref{fig:histograms}.

We evaluate the consistency of our three patches by comparing the patchwise minimum variance power spectra $C_L^p, p\in\{\mathrm{RA23,RA12,RA4.5}\}$ between pairs of patches using PTEs of the quantities $C_L^p-C_L^{p'}$. Additionally, we note that the first bin in the power spectrum for RA12 is considerably higher than the other two so we also evaluate PTEs specifically comparing the values in the first bin of each of our three patches. The results of these tests are summarized in Table~\ref{tab:patch_consistency} which confirm that the three patches are consistent with each other.

\begin{table}[htbp]
\begin{center}
\caption{\label{tab:patch_consistency} PTEs comparing pairs of observational patches.}
\begin{tabular}{lrr}
\hline
                & full spectra & first bins only \\
\hline
RA23  vs. RA12  & 24.8\% & 35.6\% \\
RA12  vs. RA4.5 & 49.6\% &  8.0\% \\
RA4.5 vs. RA23  & 89.2\% & 56.2\% \\
\hline
\end{tabular}
\end{center}
%\tablecomments{
We evaluate the consistency of our three patches with PTE values for the quantity $C_L^p-C_L^{p'}$.
%}
\end{table}

This modest excess of power in RA12 is also seen in cross correlation analyses with \herschel\ and \hsc\ \cite[]{Namikawa_2019, Faundez:2019lui}, both of which use independent analysis pipelines. In particular when looking at the \herschel\ galaxy auto-power spectra of RA23 and RA12 patches, we see that RA12 has a modest excess in power in the lowest multipole bin similar to what we see in the present analysis. This gives further support to the interpretation that the larger power in RA12 at low multipoles is due to cosmic variance.

We test the consistency of this result with the \lcdm\ cosmology using cosmological parameters from \cite[]{2016A&A...594A..13P}. We evaluate PTE values for the $\chi^2$ statistic
\beq \chi^2 = \sum_{L\alpha L'\alpha'} \left(C_{L}^{\alpha} - \Alense C_L^{\alpha\mathrm{(th)}}\right) \invcov \left(C_{L'}^{\alpha'} - \Alense C_{L'}^{\alpha'\mathrm{(th)}}\right), \eeq
which is summed over power spectra for all estimators and patches, nine in total. Relative to our best fit value of \Alens{} = \alensval\ that we achieve from the \pb\ dataset, we find a PTE of 58.8\%, and relative to the value of \Alens{} = 1 expected from \lcdm\ we find a PTE of 5.0\%, which indicate that our results agree with our current understanding of the cosmological standard model.

We can also consider our result in light of recent results for \Alens\ from Planck. The lensing smoothing effect on the Planck temperature and polarization CMB power spectra give a value of \Alens{} $ = 1.180 \pm 0.065$ \cite[]{2018arXiv180706209P} which differs from \Alens{} $=1$ by $2.8\sigma$, while the measurement from Planck lensing reconstruction \cite[]{2018arXiv180706210P} is consistent with \lcdm\ with the value \Alens{} $ = 1.011 \pm 0.028$.
While our own value is also consistent with \lcdm, we note that our estimate of \Alens\ is consistent with both of these Planck estimates of the lensing amplitude.

\section{Conclusion}

We have presented a measurement of gravitational lensing of the polarized CMB.
This work was performed using a blind analysis framework that subjected our dataset to a suite of null tests to validate error-bars and show that our data selection and analysis pipeline are not contaminated by unknown systematic errors.
We include the impact of known foreground and instrumental systematic errors in our final estimations, and we reject the no-lensing hypothesis at a significance of \nolenssig. The lensing power spectrum derived using a minimum-variance estimator from the second season data is measured as \Alens{} = \alensvalanderr, which is a \lenssig\ measurement and is consistent with the current \lcdm\ cosmology.

The lensing information in the \pb\ data presented here is derived from polarization information. Polarization measurements of gravitational lensing will become increasingly more relevant as more experiments are dominated by polarization rather than temperature information.
This work joins our other cross correlation \cite[]{Faundez:2019lui,Namikawa_2019} and delensing \cite[]{2019arXiv190913832A} analyses in exploring signals of gravitational lensing present in CMB polarization.

\section{Acknowledgements}

The \pb{} project is funded by the National Science Foundation under Grants No. AST-0618398 and No. AST-1212230.
The James Ax Observatory operates in the Parque Astron\'omico Atacama in Northern Chile under the auspices of the Comisi\'on Nacional de Investigaci\'on Cient\'ifica y Tecnol\'ogica de Chile (CONICYT).
This research used resources of the Central Computing System, owned and operated by the Computing Research Center at KEK, the HPCI system (Project ID:hp150132), the National Energy Research Scientific Computing Center, a DOE Office of Science User Facility supported by the Office of Science of the U.S. Department of Energy under Contract No.  DE-AC02-05CH11231, and the Open Science Grid, which is supported by the National Science Foundation and the U.S. Department of Energy's Office of Science.
In Japan, this work was supported by MEXT KAKENHI Grant Numbers JP15H05891, 21111002, JSPS KAKENHI Grant Numbers JP26220709, JP24111715, JP26800125, and JP18H05539.
This work was supported by World Premier International Research Center Initiative (WPI), MEXT, Japan.
This work was supported by JSPS Core-to-Core program.
In Italy, this work was supported by the RADIOFOREGROUNDS grant of the European Union’s Horizon 2020 research and innovation programme (COMPET-05-2015, grant agreement number 687312) as well as by the INDARK INFN Initiative and the COSMOS network of the Italian Space Agency (cosmosnet.it).
The McGill authors acknowledge funding from the Natural Sciences and Engineering Research Council of Canada and the Canadian Institute for Advanced Research.
Support from the Ax Center for Experimental Cosmology at UC San Diego is gratefully acknowledged.
The Melbourne group acknowledges support from the University of Melbourne and an Australian Research Council’s Future Fellowship (FT150100074).
Work at LBNL is supported in part by the U.S. Department of Energy, Office of Science, Office of High Energy Physics, under contract No. DE-AC02-05CH11231.
This work was supported by the Moore Foundation, the Templeton Foundation and the Simons Foundation.

GF acknowledges the support of the European Research Council under the European Union's Seventh Framework Programme (FP/2007-2013) / ERC Grant Agreement No. [616170] and of the UK STFC grant ST/P000525/1.
MA acknowledges support from CONICYT UC Berkeley-Chile Seed Grant (CLAS fund) Number 77047, Fondecyt project 1130777 and 1171811, DFI postgraduate scholarship program and DFI Postgraduate Competitive Fund for Support in the Attendance to Scientific Events.
MH acknowledges the support from the JSPS KAKENHI Grant Numbers JP26220709 and JP15H05891.
JC acknowledges support from the European Research Council under the European Union's Seventh Framework Programme (FP/2007-2013) / ERC Grant Agreement No. [616170]
YC acknowledges the support from the JSPS KAKENHI Grant Number 18K13558 and 18H04347.
AK acknowledges the support by JSPS Leading Initiative for Excellent Young Researchers (LEADER) and by the JSPS KAKENHI Grant Number JP16K21744.
BDS acknowledges funding from the European Research Council (ERC) under the European Union’s Horizon 2020 research and innovation programme and support from an STFC Ernest Rutherford Fellowship.
OT acknowledges the support from the JSPS KAKENHI Grant Number JP26105519.

\begin{appendix}

\section{Data Split Null Test PTE Values} \label{app:split_ptes}

As mentioned in Section~\ref{sec:analysis}, we estimate a realization-dependent noise bias as part of calculating the lensing spectrum for our dataset. This step is computationally expensive and ideally we would estimate a realization-dependent bias for the data and each of the 500 \mc\ simulations to ensure they are all treated exactly the same by our analysis pipeline. However it was only computationally reasonable for us to estimate a realization-dependent bias for the data, and instead we use the easier to calculate Gaussian bias for each of our simulations.

A possible effect of using a slightly more accurate noise bias subtraction on the data than on the simulations is that we may get higher PTEs in our null tests than if we had used a realization-dependent bias for all simulations. In Table~\ref{tab:pte_summary_split} it appears that such an effect might be resulting in high PTEs, considering the fact that out of 18 statistics the lowest PTE value is 35.8\%. To determine if these high values are the result of the difference in bias calculation between data and simulations, we perform an additional set of tests.

These new tests differ from our default pipeline in that we use the simpler to calculate Gaussian bias for both data and simulations. This will result in a less accurate calculation of the lensing spectrum for our data, but it treats all calculations equally. If the resulting PTEs from this set of null tests are lower than their counterparts in Table~\ref{tab:pte_summary_split} then we have evidence that the realization-dependent bias subtraction is the source of high PTEs in our data split null tests.
When we performed this new test the results showed that nearly all the PTE values decreased (including all 12 of the worst $\chi^2$ PTEs) as is consistent with this hypothesis.

As to the implications of this on our final results, not using a realization-dependent bias subtraction for simulations may be resulting in slightly larger error bars and a more conservative estimate of our detection significance. As an approximate estimate of how big this effect might be, we look at the distributions of \chinullb\ values for each patch and estimator to see how much the standard deviation of the distribution of these \chinullb\ changes. The effect varies for each estimator, but on average we see a 4\% change in the standard deviation of the distribution when comparing results with and without the realization-dependent calculation. This is a small effect, on the order of the uncertainty from using a finite number of simulations, and so we are confident that the results in Table~\ref{tab:pte_summary_split} still serve as an acceptable check on unknown systematic biases in the rest of our analysis pipeline.

\end{appendix}

\bibliographystyle{apj}
\bibliography{pb}

\end{document}